\documentclass[apj,twocolumn]{openjournal}
\usepackage{amsmath}
\graphicspath{{./}{figures/}}

\usepackage{xcolor}
\definecolor{xlinkcolor}{cmyk}{1.0, 0.31, 0, 0.42}

\PassOptionsToPackage{hyphens}{url}
\usepackage[bookmarks=true, pdfnewwindow=true, colorlinks=true, linkcolor=xlinkcolor, citecolor=xlinkcolor, filecolor=xlinkcolor, urlcolor=xlinkcolor, final=true]{hyperref}

\newcommand{\orcidauthor}[3]{\author{\href{http://orcid.org/#1}{#2$^{#3}$}}}

\shorttitle{MultiNest: Maladies and Remedies}
\shortauthors{A. J. Dittmann}

\begin{document}

\title{\vspace{-0.8cm}Notes on the Practical Application of Nested Sampling: \\  MultiNest, (non)convergence, and rectification\vspace{-1.8cm}}

\orcidauthor{0000-0001-6157-6722}{Alexander J. Dittmann}{1,*}
\thanks{$^*$\href{mailto:dittmann@umd.edu}{dittmann@umd.edu}}
\affiliation{$^{1}$Department of Astronomy and Joint Space-Science Institute, University of Maryland, College Park, MD 20742-2421, USA}

\begin{abstract}
Nested sampling is a promising tool for Bayesian statistical analysis because it simultaneously performs parameter estimation and facilitates model comparison. MultiNest is one of the most popular nested sampling implementations, and has been applied to a wide variety of problems in the physical sciences. 
However, MultiNest results, like those of any sampling tool, can be unreliable, and accompanying convergence tests are a \textit{necessary} component of any analysis.
Using analytically tractable test problems, I illustrate how MultiNest, when applied without rigorously chosen hyperparameters,
(1) can produce systematically erroneous estimates of the Bayesian evidence, which are more significantly biased for problems of higher dimensionality; 
(2) can derive posterior estimates with errors on the order of $\sim100\%$;  
(3) can, particularly when sampling noisy likelihood functions, systematically underestimate posterior widths.
Furthermore, I show how MultiNest,  thanks to the advantageous speed at which it explores parameter space, can  also be used to jump-start Markov chain Monte Carlo sampling or more rigorous nested sampling techniques, potentially accelerating more  robust measurements of posterior distributions and Bayesian evidences, and overcoming the challenge of Markov chain Monte Carlo initialization.   
\end{abstract}


\section{Introduction}\label{sec:intro}
Fitting models to data and comparing the ability of different models to describe observations are integral to scientific practice. 
Nested sampling \citep{2004AIPC..735..395S} is a technique that can estimate both posteriors and Bayesian evidences, which has led to its widespread popularity in the physical sciences \citep[see][for recent reviews]{2022NRvMP...2...39A,2023StSur..17..169B}. 

MultiNest \citep{2008MNRAS.384..449F,2009MNRAS.398.1601F} is arguably the most popular nested sampling implementation. It is relatively fast (see, for example, Figure 7 of \citet{2015MNRAS.450L..61H} or Figure 4 of \citet{2020arXiv201215286A}), and it can provide accurate results 
 when used with sufficiently rigorous settings.
MultiNest has been applied to X-ray astronomy \citep[e.g.,][]{2014A&A...564A.125B,2019ApJ...887L..24M,2019ApJ...887L..21R}, the study of exoplanets \citep[e.g.][]{2013MNRAS.434L..51K,2016ApJ...822...86M,2018AJ....155..156T}, and cosmology \citep[e.g.][]{2014A&A...571A..22P,2016A&A...594A..20P,2020A&A...641A..10P}, in addition to many other fields. 

However, numerous studies have found that MultiNest can perform poorly and unreliably, both in terms of inferring posterior distributions \citep[e.g.,][]{2019ApJ...887L..24M,2021AJ....162..237I,2021ApJ...918L..28M} and measuring Bayesian evidences \citep[e.g.,][]{2020arXiv201215286A,2020AJ....159...73N,2023MNRAS.521.1184L}. One of the challenges faced when using MultiNest,  like virtually all Bayesian analysis tools, is deciding on values for its free parameters. The most important of these parameters are the number of live points ($N_{\rm live},$ which determines how densely parameter space is sampled) and the sampling efficiency ($\epsilon$, which controls how conservatively the algorithm explores parameter space). One can also tune the algorithm's stopping criterion through a tolerance parameter for the evidence (\texttt{tol}), but that is usually of little consequence to the algorithm's performance.\footnote{Unless one's goal is maximum-likelihood estimation: As MultiNest marches higher and higher toward the peak of the likelihood function, eventually the remaining prior volume is so minuscule that the remainder of the evidence integral is negligible, and the algorithm terminates. Setting a smaller tolerance value continues iteration until less and less of the evidence integral remains.  Regardless, the small number of free parameters in the MultiNest algorithm are an attractive feature, making testing its convergence a straightforward process.}

Unfortunately, the values for these parameters  suggested by the MultiNest papers and documentation, upon which many users rely, are often inadequate (see Section \ref{sec:problems}). For example, the \texttt{README} file included with the MultiNest source code suggests using $\epsilon=0.8$ for parameter inference and  $\epsilon=0.3$ for evidence measurement.\footnote{\url{https://github.com/farhanferoz/MultiNest/blob/cc616c17ed440fbff8d35937997c86256ed4d744/README.md}} The primary MultiNest paper states that  ``we found 400 and 50 active points to be adequate for evidence evaluation and parameter estimation respectively,'' and suggests sampling efficiency values of $1$ and $0.3$ for parameter estimation and evidence evaluation \citep[][Section 8]{2009MNRAS.398.1601F}. As shown in Section \ref{sec:problems}, these suggestions about sampling posteriors are reasonable for simple Gaussian distributions. However, even for ten dimensional Gaussian distributions these suggested parameter values are inadequate to accurately measure the Bayesian evidence.\footnote{It is worth noting that the authors of MultiNest have advocated that ``it is always advisable to increase $N_{\rm live},$ and reduce [$\epsilon$]
to check if the posteriors and evidence values are stable," \citep{2019OJAp....2E..10F} although many papers have used MultiNest without reporting such convergence tests.}

Many studies have investigated potential deficiencies in nested sampling algorithms, developed algorithmic improvements, and suggested methods for testing convergence. To give a few examples, \citet{2020AJ....159...73N,2020arXiv201215286A,2023MNRAS.521.1184L} have noted systematic, and sometimes dimension- and sampling efficiency-dependent biases in the evidence estimates produced by MultiNest. \citet{2016S&C....26..383B} developed a test that illustrated MultiNest's tendency to sample from an erroneously small prior volume, and developed an alternative algorithm for unbiased, albeit slower, sampling. \citet{2019MNRAS.483.2044H} proposed some strategies to check convergence by cross-checking multiple analyses, and advocated that when using MultiNest one should decrease $\epsilon$ and increase $N_{\rm live}$ to check convergence.  

My aims here are to systematically study MultiNest's convergence in terms of both the sampling efficiency and number of live points for problems of varying dimensionality, and to present strategies that can be applied to improve upon completed nested sampling analyses , taking advantage of MultiNest's favorable speed. Section \ref{sec:nsmn} provides a brief overview of MultiNest and nested sampling, after which Section \ref{sec:methods} describes some of my methodology. The results of my tests using MultiNest to sample smooth distributions are presented in Section \ref{sec:problems}, and my exploration of MultiNest's performance on a noisy target distribution is presented in Section \ref{section:noisy}. Section \ref{sec:improve} demonstrates a couple of methods that can be used to derive better posterior and evidence estimates, using an initial MultiNest (or other nested sampling) run to accelerate the process. 

The scripts used to run each of these tests are hosted on GitHub.\footnote{\url{https://github.com/ajdittmann/multiNestNotes}} This work was conducted using MultiNest's python bindings \citep{2016ascl.soft06005B} and MultiNest version 3.12.\footnote{\url{https://github.com/farhanferoz/MultiNest}}

\section{The MultiNest Algorithm}\label{sec:nsmn}
Nested sampling begins by drawing a set of $N_{\rm live}$ points from the prior. Each iteration of the algorithm proceeds by replacing the lowest-likelihood point with a new higher-likelihood point. If the higher-likelihood point is drawn uniformly from within the isolikelihood surface of the lowest-likelihood point, then the expectation value of the prior volume encompassed by the set of live points at iteration $i$ will be $\exp{(-i/N_{\rm live})}$.\footnote{Notably, this estimate, which I focus on here because of its usage in MultiNest \citep{2009MNRAS.398.1601F}, is very slightly biased \citep[e.g.,][]{Walter2017,2018arXiv180503924S,2023StSur..17..169B}. } In this way, each sample can be associated with a fraction of the prior volume. The product of each point's likelihood and associated prior volume then yields its posterior weight, and it is straightforward to integrate the likelihood over the prior to approximate the evidence.

If one were to perform simple rejection sampling, attempting to draw samples from the entire prior, the odds of drawing a point with a higher likelihood than those in the set of live points would become (exponentially) vanishing small. MultiNest's innovative approach was to, rather than sample from the full prior, first fit a hyperellipsoid (or set of hyperellipsoids, if favored by a k-means algorithm) to the set of live points, to potentially expand those ellipsoids by a factor $f$, and to sample from within the resulting set of ellipsoids. At each iteration, the factor $f$ is chosen such that $f={\rm max}[1,\exp{(-i/N_{\rm live})}V(E)^{-1}\epsilon^{-1})]$, where $\epsilon$ is the sampling efficiency parameter and $V(E)$ is the volume contained within minimum bounding ellipsoids. That is to say, \textit{if} the volume of the ellipsoids bounding the set of live points is smaller than the expectation value of the remaining prior volume divided by the sampling efficiency, the ellipsoids are expanded so that their volume matches that target. See Section 5 of \citet{2009MNRAS.398.1601F} for a more thorough presentation.

The sampling efficiency factor $\epsilon$ acts as a safeguard in the algorithm, making it more likely that new points are actually drawn from within the intended likelihood contour. However, when using $\epsilon=0.3$, for example, MultiNest is still wont to sample from too small a volume of parameter space \citep{2016S&C....26..383B}. Notably, the amount of ellipsoid expansion in any given linear dimension at constant $\epsilon$ is much smaller in higher dimensions. This is why smaller efficiencies are required to produce accurate results in higher dimensions and why using sampling efficiencies greater than unity is much more damaging for lower-dimension problems, as shown in Section \ref{sec:problems}. 

\section{Test Problems and Methodology}\label{sec:methods}
I have chosen three problems to test MultiNest's robustness when applied to smooth problems: multidimensional versions of 
the normal distribution, the Rosenbrock distribution, and the log-Gamma distribution.
In each test the integral of the target distribution can be calculated analytically, and after normalizing the log likelihood the evidence is simply given by the prior normalization. 

The one-dimensional posterior distributions in the normal and log-Gamma distribution tests are trivial to calculate, so one can easily test the ability of each sampler to correctly recover the underlying posterior distribution. I took a practical approach to estimating errors in posterior recovery: I first computed a kernel-density estimate using the posterior samples drawn in each test  (between $\sim 2000$ and $\sim530000$ samples depending predominantly on the problem dimensionality and number of live points), and then calculated the $L_1$ error of that kernel density estimate and the true underlying distribution. Explicitly, I calculated in each dimension ($i$)
\begin{equation}\label{eq:l1}
L_{1,i} = \int \left| p(x_i) - \tilde{p}(x_i)\right| dx_i,
\end{equation}
where $p(x_i)$ is the true marginalized posterior distribution and $\tilde{p}(x_i)$ is a kernel density estimate derived from the posterior samples. I used the scipy \citep{2020SciPy-NMeth} implementation of the QUADPACK routine \citep{1983qspa.book.....P} to evaluate the integral (\ref{eq:l1}), and the default scipy Gaussian KDE implementation, which used Scott's rule to select the bandwidth \citep{2015mdet.book.....S}. When reporting the results of various tests, I have presented the distribution of $L_1$ errors over all dimensions and independent trials. 

In addition to the overall error in a given analysis, the deviations between the true posteriors and a set of posteriors estimates, it is useful to know if an algorithm tends to overestimate or underestimate the widths of credible intervals. To test this, in some cases I have calculated the \% error of the $\pm 1\sigma$ ($\approx 68\%$) credible region, defined such that a negative $\%$ error means that MultiNest underestimated the width of a given credible interval.  These estimates have the benefit of being derived directly from the posterior samples, without any binning or density estimation.

Because each of the likelihood functions examined here can be integrated analytically, I have calculated evidence errors according to 
\begin{equation}
\Delta \log{Z}=\log{Z_{\rm measured}}-\log{Z_{\rm true}},
\end{equation}
such that positive values of $\Delta\log{Z}$ indicate that MultiNest overestimated the evidence.  This sort of evidence error analysis has been conducted previously in \citet{Brewer2011,2020arXiv201215286A,2020AJ....159...73N,2023MNRAS.521.1184L}, for example.

\subsection{Test Problems}
\subsubsection{The Normal Distribution}\label{sec:normal}
The simplest distribution I have considered here is a multivariate uncorrelated normal distribution.
In general, most local maxima of realistic log likelihoods can be approximated by parabolas, so normal distributions will be relevant in at least some neighborhood about each log likelihood peak. However, in practice, many realistic log likelihoods deviate appreciably. 
The log-likelihood function in $N$ dimensions is given by
\begin{equation}
\log L(\mathbf{x}) = \sum_i -\frac{1}{2}\left(\frac{x_i^2}{\sigma_i}\right)^2 - \log{\sigma_i\sqrt{2\pi}}.
\end{equation}
For each problem, I selected standard deviations $\sigma_i$ uniformly between 0.1 and 0.5 (inclusive), with a spacing dependent on the number of dimensions. 
For this problem, I used uniform priors in each dimension on the interval $[-20,20]$. Because this problem is unimodal with a likelihood gradient only increasing further from its peak, the entire prior serving as an effective basin of attraction, the prior widths have negligible influence, and do not affect any of the conclusions drawn from these tests.\footnote{A quick test one can run to verify this is sampling a 20-dimensional normal distribution, as described here, using $N_{\rm live}=300$ and $\epsilon=0.3$. As an example, using the default priors adopted here, ten independent analyses had a mean log evidence error of $\Delta\log{Z}=1.35$ with a standard deviation of $0.26$. Shrinking the priors to $[-5,5]$ in each dimension and running again, I found a mean error of $\Delta\log{Z}=1.20$ with a standard deviation of $0.31$. Similar results hold with more live points (see section \ref{sec:normals}) or even tighter priors.}

\subsubsection{An Extended Rosenbrock Distribution}\label{sec:rosenbrock}
The Rosenbrock function in two dimensions is a standard test problem in numerical optimization, notable for its long, curving degeneracies, and strong correlations between parameters. There have been many attempts to extend the Rosenbrock function to higher dimensions, but I focus here only on the distribution presented in \citet{2019arXiv190309556P}, which exhibits nontrivial correlations between each pair of parameters and has a known normalization. 

The log likelihood for this distribution is given by, in $N=(n_1-1)n_2+1$ dimensions,
\begin{equation}
\log L(\mathbf{x}) = -a(x_1-\mu)^2 - \sum_{j=1}^{n_2}\sum_{i=2}^{n_1}b_{j,i}(x_{j,i}-x^2_{j,i-1})^2-C,
\end{equation}
where the normalization constant $C$ is given by 
\begin{equation}
C=\log{\left(\frac{\sqrt{a}\prod\sqrt{b_{j,i}}}{\pi^{N/2}}\right)}.
\end{equation}
I chose $\mu=0$, $a=2.0$, and each $b_{j,i}$ was chosen between 1 and 10 (inclusive) with a geometrically uniform spacing depending on the total problem dimensionality. For this problem, I used uniform priors in each dimension on the interval $[-50,50]$. As with the normal distribution, the priors are largely irrelevant, provided they are wide enough for the analytical evidence value to apply.

\subsubsection{The Log-Gamma Distribution}\label{sec:logg}
Many distributions possess heavy tails.
It is well-known that such distributions pose challenges to MultiNest, although the standard test involves four separate modes and combinations of log-gamma and normal distributions \citep[e.g.,][]{2013arXiv1304.7808B,2016S&C....26..383B,2019OJAp....2E..10F}. 
I have instead chosen a unimodal variation of this problem without any Gaussian components. The log likelihood for this test is given by 
\begin{align}
\log L(\mathbf{x}) = \sum_i cy_i-\exp{y_i} - \log{\Gamma(c)},\\
y_i = (x_i - \mu)/\sigma_i,
\end{align}
where $\Gamma$ is the Gamma function, $c=1$, $\mu=2/3$, and scale parameters $\sigma_i$ were spaced log-uniformly between 0.01 and 0.05 (inclusive). For this problem, I used uniform priors in each dimension on the interval $[-10,2]$, and unlike the other tests, in this case priors can have a substantial effect. As discussed in Section \ref{sec:logGammaTests}, MultiNest has a tendency to miss the peak of this distribution and instead hone in on some portion of its long tail in various numbers of dimensions. More narrow priors, particularly on the lower end, would limit the range over which MultiNest could get stuck; wider or narrower priors can thus lead to greater or lower magnitude evidence errors, although the priors have little-to-no effect on which hyperparameters are necessary to thoroughly sample the distribution.

\subsubsection{A Noisy Normal Distribution}\label{sec:noisygauss}
While attempting to understand why MultiNest might perform fairly reliable posterior inference for many standard test problems 
but in certain astronomical data analyses systematically underestimate credible region widths
\citep[e.g.,][]{2019ApJ...887L..24M,2021AJ....162..237I,2021ApJ...918L..28M},\footnote{Most quantitatively, in \citet{2021ApJ...918L..28M}, MultiNest underestimated the neutron star equatorial radius posterior width by a factor of $\sim0.67$ when used with ($N_{\rm live}=1000,$ $\epsilon=0.01$), a factor of $\sim0.79$ when used with ($N_{\rm live}=3000,$ $\epsilon=0.01$), and even smaller factors when used with larger sampling efficiencies.} I considered the admittedly speculative possibility that some sort of ``noise'' (perhaps some fairly high-frequency or stochastic variation in the likelihood as a function of one or more parameters, superimposed over some smooth trend) might be present in actual analyses but not the smooth test problems described above.\footnote{ I was led to this line of investigation by first using a KDE of the posterior samples from \citet{2021ApJ...918L..28M} as a target likelihood function and finding that MultiNest produced accurate posterior inferences (for a range of KDE bandwidths), suggesting that some property of the analysis besides simply the \textit{shape} of the posteriors was to blame.} To test this idea, I used the following log-likelihood function in N dimensions
\begin{equation}\label{eq:noisygauss}
\log L(\mathbf{x}) = \sum_i -\frac{1}{2}\left(\frac{x_i^2}{\sigma_i}\right)^2 - Af(x_i),
\end{equation}
where $f(x)$ is a linear interpolant over a sequence of pseudorandom numbers drawn uniformly between -1 and 1, and $A$ is a noise amplitude. For simplicity I used the same pseudorandom field in each dimension, which had between 250 and 1250 samples per $\sigma_i$. The standard deviations and priors were the same as in Section \ref{sec:normal}. To be clear, I make no claim that this sort of noise is actually present in practical analyses: this is simply one test case where MultiNest performs similarly to the aforementioned applications to real data. 
 \begin{figure*}
 \includegraphics[width=\linewidth]{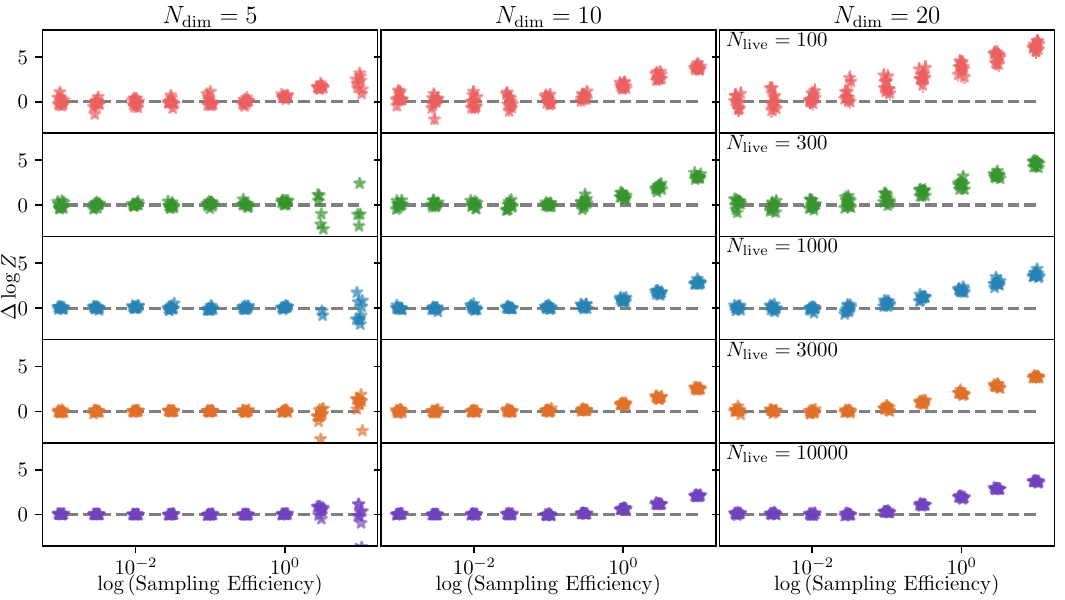}
 \caption{Errors in MultiNest-derived evidence values from the multivariate normal distribution tests. Each column plots results from distributions of different dimensionalities, each row plots results from tests using different numbers of live points, and results in each subplot are plotted as functions of sampling efficiency. Significant departures from the gray line, which marks an evidence error of zero, indicate cases in which MultiNest failed to measure the evidence accurately. Error bars indicate the MultiNest-estimated uncertainties, although these are often dwarfed by the actual errors. }
 \label{fig:gaussEv}
 \vspace*{-0.5mm}
\end{figure*}

\begin{figure*}
 \includegraphics[width=\linewidth]{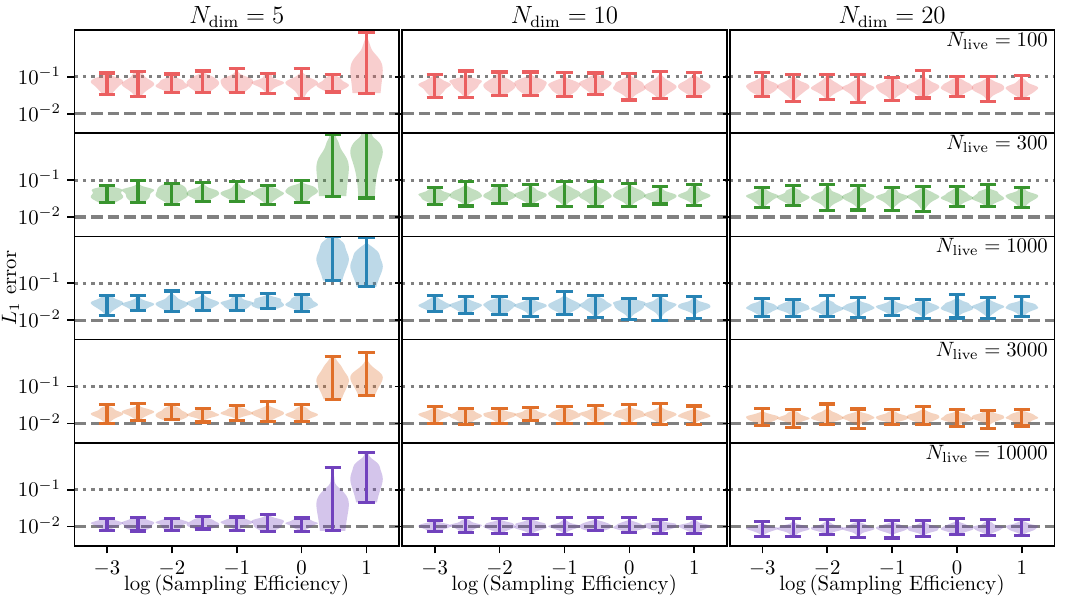}
 \caption{Errors in MultiNest-derived posteriors from the multivariate normal distribution tests. The layout is the same as in Figure \ref{fig:gaussEv}. Dashed and dotted gray lines mark $L_1$ errors of 0.01 and 0.1 respectively.}
  \vspace*{-0.0mm}
 \label{fig:gaussPost}
\end{figure*}

\begin{figure*}
 \includegraphics[width=\linewidth]{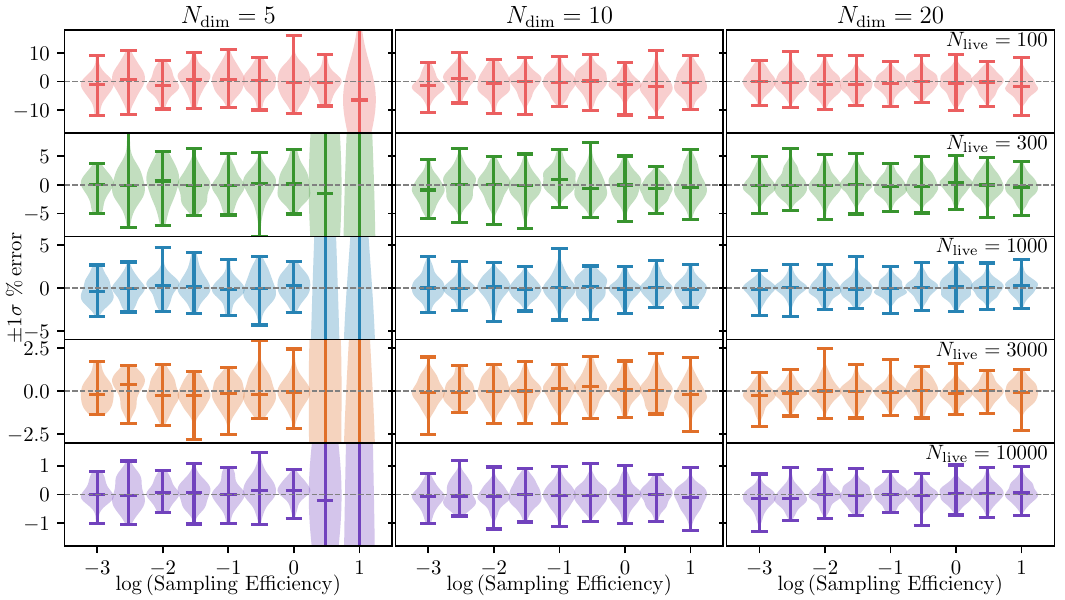}
 \caption{Errors in the widths of the MultiNest-derived $\pm1\sigma$ credible regions from the multivariate normal distribution tests.  The horizontal lines in each violin plot denote the minimum, median, and maximum values.}
  \vspace*{-3.5mm}
 \label{fig:gaussWidths}
\end{figure*}
\begin{figure*}

 \includegraphics[width=\linewidth]{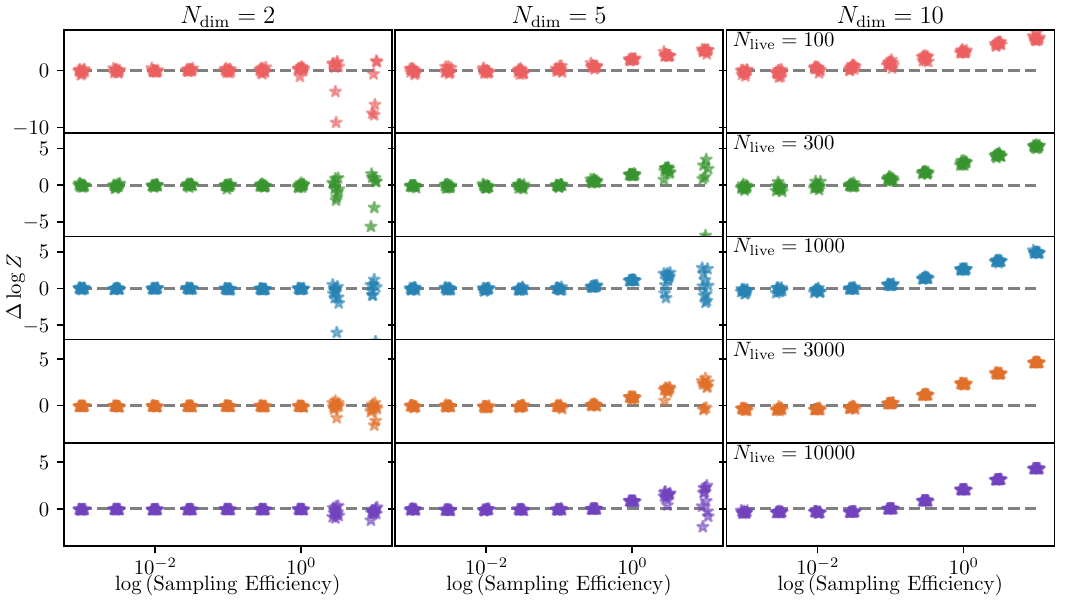}
 \caption{Errors in MultiNest-derived evidence values from the generalized Rosenbrock distribution tests. Each column plots results from distributions of different dimensionalities, each row plots results from tests using different numbers of live points, and results in each subplot are plotted as functions of sampling efficiency. Error bars again indicate the MultiNest-estimated uncertainties, although these are frequently orders of magnitude smaller than the actual errors. }
 \label{fig:banana}
\end{figure*}

\subsection{Testing Protocols}
\subsubsection{Smooth Test Problems}
For each smooth toy problem (those described in Sections \ref{sec:normal}, \ref{sec:rosenbrock}, and \ref{sec:logg}, and analyzed in Section \ref{sec:problems}), I have tested a wide variety of resolution parameters to gauge the MultiNest's performance. Concretely, I used $N_{\rm live}\in\{100,300,1000,3000,10000\}$, for each $N_{\rm live}$ I used the sampling efficiencies $\epsilon~\in~\{10, 3, 1, 0.3, 0.1, 0.03, 0.01, 0.003, 0.001\}$, and I tested each ($N_{\rm live}$, $\epsilon$) combination 10 independent times (changing only the pseudorandom seed) for each of the three test problems. For each problem, I have also investigated three different dimensionalities (e.g. 5D, 10D, and 20D for the multidimensional normal distribution test), totaling 4050 analyses. 

I have focused on only the standard MultiNest algorithm \citep[][version 3.12]{2009MNRAS.398.1601F}, without using importance nested sampling \citep{2019OJAp....2E..10F}. Since this variation does not alter MultiNest's parameter exploration strategy, it would not ameliorate any of the cases where MultiNest's posterior estimates erred. Importance nested sampling might improve performance in some of the cases where the posterior estimates were acceptable and the evidence was simply overestimated, although it is far from panacean \citep[e.g.,][]{2020AJ....159...73N}, sometimes leading to the confident exclusion of the true evidence \citep{2016S&C....26..383B}.

Although each problem was unimodal, I allowed MultiNest to perform mode separation if it identified multiple modes, since in practice one rarely knows a priori how many modes are present in a given problem. I performed a handful of log-gamma tests with mode separation deactivated and observed no significant changes in the results. I set MultiNest's evidence tolerance parameter to $10^{-5}$ in each analysis to avoid the possibility that parameter space might not be thoroughly explored because the tolerance was too large, although I also repeated $\sim100$ normal and log-gamma tests using a tolerance parameter of $0.25$ and found no appreciable differences in the quality of the results.  Because each problem is unimodal and none are characterized by log-likelihood plateaus, MultiNest's performance is largely insensitive to the prior bounds, as described above. 
However, if multiple modes may be present, as is generally the case, one must be careful to use enough live points for MultiNest to have a good chance of finding each relevant mode within the prior volume \citep[see, for example the discussion in Section 9 of][]{2008MNRAS.384..449F}.

\subsubsection{A Noisy Test Problem}
For the test introduced in Section \ref{sec:noisygauss}, developed in an attempt to understand a possible cause for MultiNest systematically underestimating credible region widths \citep[e.g.,][]{2019ApJ...887L..24M,2021AJ....162..237I} in a sampling-efficiency-dependent way \citep[e.g.,][]{2021ApJ...918L..28M}, I focused instead on MultiNest parameters closer to those used in scientific applications. 
Specifically, I used $N_{\rm live}\in\{300,1000,3000,10000\}$, for each $N_{\rm live}$ I have used the sampling efficiencies $\epsilon~\in~\{0.3, 0.1, 0.03, 0.01\}$, and tested each ($N_{\rm live}$, $\epsilon$) combination 10 independent times. For each of these, I have saved MultiNest outputs at two evidence tolerance values, $\texttt{tol}\in\{0.5,0.01\}$, waiting for the first series to complete and then re-starting each with a lower tolerance. For succinctness, I have only reported in Section \ref{section:noisy} results from a 10D test case. 

\section{MultiNest in Action:\\ Smooth Test Problems}\label{sec:problems}
\subsection{Normal Distributions}\label{sec:normals}
\begin{itemize}
\item When using too large a sampling efficiency, MultiNest overestimates the evidence, more so in higher dimensions.
\item For evidence measurement in lower-dimension (5) problems, $\epsilon\lesssim 0.3$ may be sufficient at $N_{\rm live}=10^2$, and $\epsilon\lesssim 1$ may be sufficient at $N_{\rm live}=10^4$.
\item For higher-dimension (20) problems, $\epsilon\lesssim 0.01$ may be sufficient at $N_{\rm live}=10^2$, and $\epsilon\lesssim 0.03$ may be sufficient at $N_{\rm live}=10^4$.
\item Posterior accuracy is effectively independent of $\epsilon$ (for $\epsilon\leq1$).
\item Posterior accuracy improves at close to, but slightly more slowly than, $\propto N_{\rm live}^{-1/2}$.
\end{itemize}

\begin{figure}
 \includegraphics[width=\linewidth]{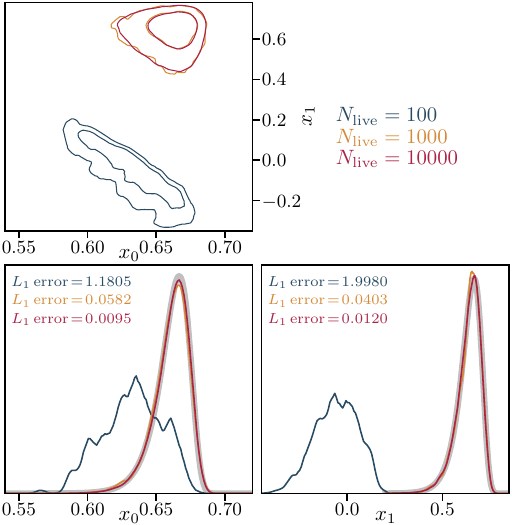}
 \caption{Posterior distributions for a 2D log-gamma distribution test using $\epsilon=1$. Blue, orange, and red curves plot results derived using $10^2$, $10^3$, and $10^4$ live points, respectively. The true one-dimensional marginalized posteriors are plotted using solid gray lines. These highlight a run that largely missed the mark, resulting in an $L_1$ error greater than 1, and two that sampled accurately to varying degrees of precision. The contours in the upper panel display the two-dimensional $1\sigma$ and $2\sigma$ minimum-area credible regions. The distributions are plotted here using averaged shifted histograms \citep{scott1985}. 
 }
 \label{fig:gammapost}
\end{figure}

These observations are based on the results presented in Figures \ref{fig:gaussEv} and \ref{fig:gaussPost}. The former figure presents the individual evidence estimates from each multivariate normal analysis, along with the uncertainties reported by MultiNest, which are plotted as error bars. In many cases the reported uncertainties are orders of magnitude smaller than the actual error of the results. The later figure plots the distribution of $L_1$ errors of the 1D posteriors in each analysis.

Figure \ref{fig:gaussWidths} displays the errors in the width of the $\pm1\sigma$ credible intervals for each of the normal distribution analyses, collated similarly to the $L_1$ error measurements. For this simple problem, errors in the posterior widths derived by MultiNest behave similarly to their $L_1$ errors, shrinking roughly according to $N_{\rm live}^{-1/2}$, and are largely independent of the sampling efficiency provided that $\epsilon\leq1$; however as shown in Section \ref{section:noisy}, this independence is not general.

 In this test problem isolikelihood surfaces can be perfectly described by hyperellipsoids, which should minimize any errors that could potentially follow from the specific choice of using ellipsoids to construct bounding regions for drawing new points, which should closely match the isolikelihood surface corresponding to the lowest-likelihood live point in order to prevent errors. However, depending on the precise distributions of live points, each bounding ellipsoid can still fail to encompass the intended isolikelihood surface \citep[see, for example, Figure 1 of][]{2023MNRAS.521.1184L}, so that samples may be drawn from an artificially small region of the prior volume at each stage. Naturally, this potential problem becomes less significant when more live points and lower sampling efficiencies are used.\footnote{ A similar bias can affect slice samplers if too few random walk iterations are used and freshly drawn points are correlated with previous points.}  Reassuringly, these errors seem to predominantly affect evidence measurements, and in most cases the posterior errors were independent of the sampling efficiency. The only cases for this problem where MultiNest produced anomalously erroneous posteriors were 5D tests using sampling efficiencies greater than unity. In these cases, MultiNest was forced to draw samples from within an artificially small hypervolume; although the same is true for the higher-dimensionality test with the same efficiency settings, in higher dimensions the shrinkage and expansion in any given dimension is diluted.

\begin{figure*}
 \includegraphics[width=\linewidth]{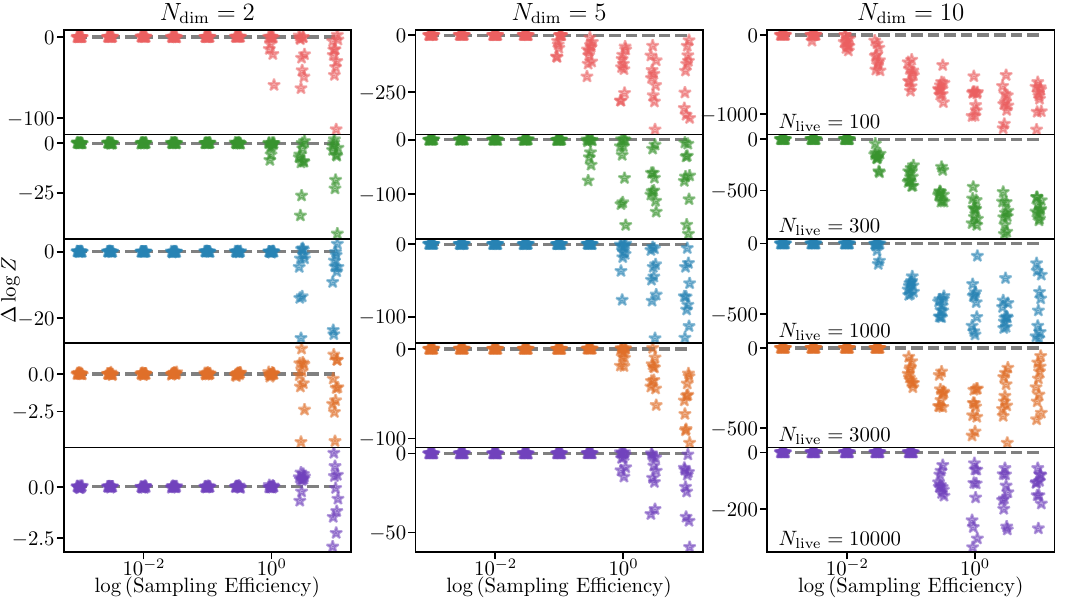}
 \caption{MultiNest evidence errors for the log-gamma problem, as functions of dimensionality, sampling efficiency, and the number of live points. Especially in higher numbers of dimensions, small sampling efficiencies and large number of live points are necessary for MultiNest to derive accurate evidence estimates.
 }
 \label{fig:gammaEv}
\end{figure*}

\begin{figure*}
 \includegraphics[width=\linewidth]{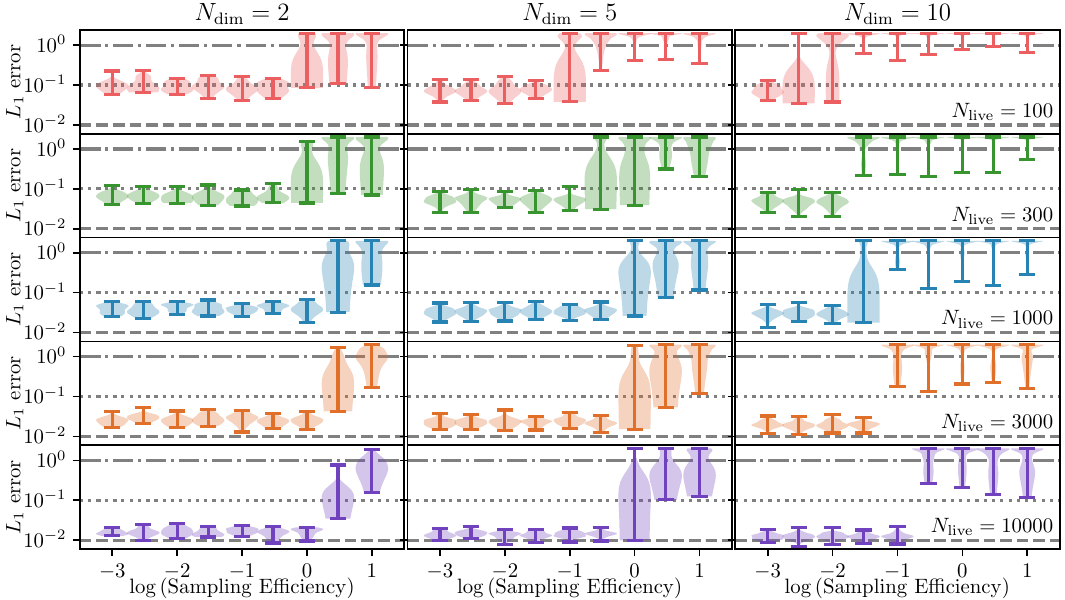}
 \caption{$L_1$ errors in MultiNest-derived posteriors from my multivariate log-gamma tests. Especially in higher numbers of dimensions, small sampling efficiencies and large number of live points are necessary for MultiNest to derive posterior estimates with that are not in complete error. }
 \label{fig:gammaErr}
\end{figure*}

\subsection{Abnormal Distributions I: Bananas \hspace{1cm} (Generalized Rosenbrock Distributions)}\label{sec:bananas}
\begin{itemize}
\item MultiNest is able to adequately sample the standard 2D Rosenbrock distribution with efficiencies of $\epsilon\leq0.3-1$ depending on the number of live points.
\end{itemize}
\begin{itemize}
\item In the 10-dimensional case, MultiNest struggles to produce accurate results more so than when sampling simpler normal distributions, requiring $\epsilon \leq 0.01$ at $N_{\rm live}\sim 100$ and $\epsilon \leq 0.1$ for $N_{\rm live}\sim 10^4$.
\end{itemize}

The generalized Rosenbrock distribution evidence estimates are presented in Figure \ref{fig:banana}. Overall, I have found that MultiNest is able to sample from generalized Rosenbrock distributions without too much additional difficulty compared to an uncorrelated multivariate normal distributions. Because the isolikelihood surfaces deviate from hyperellipsoinds, for a given dimensionality smaller sampling efficiency values are required to achieve converged results, in comparison to sampling simpler normal distributions. 

\subsection{Abnormal Distributions II: Log-Gammas}\label{sec:logGammaTests}
\begin{itemize}
\item  For the 2-dimensional problem, efficiencies between 0.3 and 1 may be sufficient to derive reasonable posteriors and estimates of the evidence, depending on the number of live points (see Figure \ref{fig:gammapost}).
\item For higher-dimension problems, efficiencies between 0.001 and 0.1 can be necessary to derive posterior and evidence estimates that are not in severe error.
\item Higher-dimension problems require lower efficiencies at a fixed number of live points.
\item Errors in MultiNest's log evidence estimates can range from 100s to 1000s when applying the developer-suggested efficiency settings to this problem.
\end{itemize}

The above observations are largely drawn from Figures \ref{fig:gammaEv} and \ref{fig:gammaErr}. The former figure presents the individual evidence estimates from each multivariate log-gamma analysis; although error bars are technically included, the estimated errors are less than unity and completely dwarfed by the systematic errors present in most of the analyses. In most of the cases where MultiNest does a very poor job of estimating the evidence, it also derives very incorrect posterior estimates, reporting results with errors on the order of $100\%$. \textit{If} this holds in general, consistent evidence measurements might suggest fairly trustworthy posterior inferences. However, as shown in Section \ref{sec:normals}, accurate posteriors \textit{do not} imply accurate evidence measurements.

A more worrying aspect of Figures \ref{fig:gammaEv} and \ref{fig:gammaErr} is the breadth over which MultiNest can report $\sim100\%$ erroneous results. Let us take the 10-dimensional problem as an example. Based on the recommendations in the MultiNest documentation, one might seemingly conservatively start with 1000 live points and an efficiency of 0.3, which would lead to a log-evidence error on the order of $\sim500.$ One may need to decrease the sampling efficiency by a factor of $\sim 300$ holding the number of live points fixed, or by a factor of $\sim 3-100$ while increasing the number of live points by factors of 10-3 to derive accurate results. However, one can not know this a priori, so diligent convergence testing should always be practiced, until the evidence estimates and posteriors converge. 

As in Section \ref{sec:bananas}, MultiNest was somewhat challenged by isolikelihood surfaces that strongly deviate from hyperellipsoids, as illustrated in Figure \ref{fig:gammapost} for the two-dimensional case. Over a wide range of hyperparameter settings, MultiNest derived posteriors with order-unity errors, the cause of which is illustrated in Figure \ref{fig:gammapost}: when too few live points or too large a sampling efficiency are used, MultiNest often misses the peak of the likelihood distribution entirely. Because the tail of the log-gamma distribution is much more shallow than the Gaussian or Rosenbrock distributions, it is much easier for all of the live points to end up in the tail with none sampling the peak. Clearly, using more live points makes it less likely to miss the peak entirely, and a smaller sampling efficiency safeguards against such situations.

\section{MultiNest in Action:\\ Sampling Noisy Likelihoods}\label{section:noisy}
\begin{figure*}
 \includegraphics[width=0.995\linewidth]{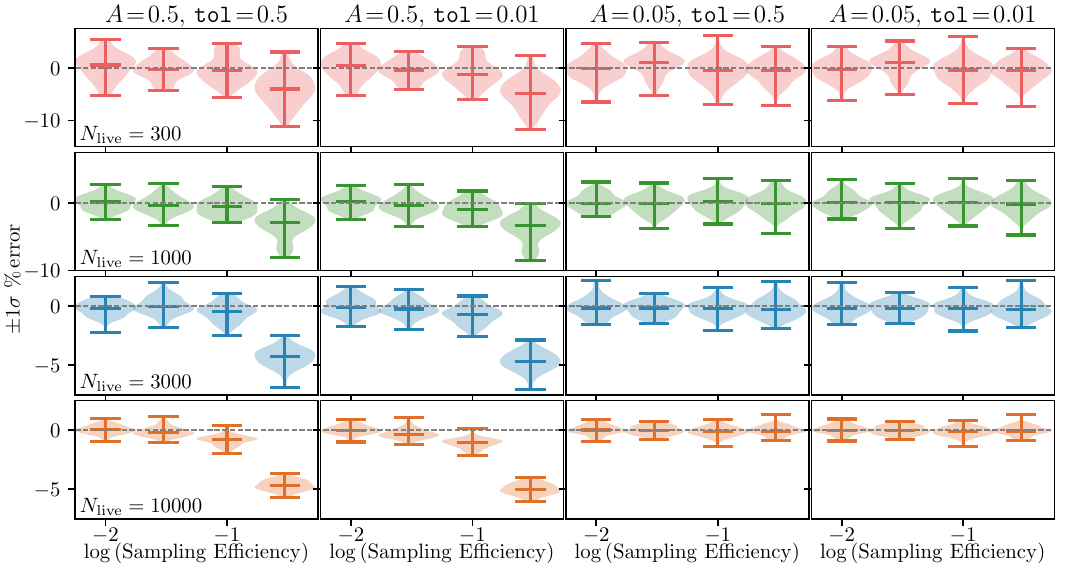}
 \caption{Errors in the widths of the MultiNest-derived $\pm1\sigma$ credible regions from the noisy normal distribution test. The horizontal lines in each violin plot denote the minimum, median, and maximum values, and the gray dotted line indicates an error of zero. At larger noise amplitudes $A$ and higher sampling efficiency values (e.g. $\epsilon=0.3$), MultiNest produces posterior estimates that are systematically narrow (by $\sim5-10\%$ for this problem). These biases remain as the number of live points is increased, but can be ameliorated by a sufficiently small value of the sampling efficiency.}
 \label{fig:spikeyErr}
\end{figure*}

Naturally, as the noise amplitude $A$ in Equation (\ref{eq:noisygauss}) goes to zero, the results should approach those of 
Section \ref{sec:normals}. I tested $A\in\{0.5,0.05, 0.005\}$, although for this test the differences between $A=0.05$ and $A=0.005$ were negligible and predictable. The results from these tests are collected in Figure \ref{fig:spikeyErr}, which shows the errors in credible region widths as functions of $N_{\rm live}$, $\epsilon$, $A,$ and \texttt{tol}. When $A$ is small, the results are effectively unbiased, as expected. However, when the noise amplitude is large, MultiNest's posterior estimates can be systematically biased towards being erroneously narrow. 

More specifically, at larger values of the sampling efficiency (such as $\epsilon=0.3,$ which is still less than half that recommended by the MultiNest documentation for parameter estimation), MultiNest produces systematically narrow posterior estimates. These biases do not disappear as more live points are used in the analyses; rather, as live points are added the run-to-run variation seems to decrease while the median error remains roughly constant between runs. The biases are significantly reduced at $\epsilon=0.1,$ and virtually disappear for this problem by $\epsilon\leq0.03$. Intriguingly, for this particular problem, results can become slightly more biased as \texttt{tol}, the evidence tolerance used by MultiNest as a stopping condition, is decreased. Presumably, because the actual likelihood surface is nonsmooth, MultiNest's bounding ellipsoids have a tendency to systematically underestimate the actual isolikelihood surface at a given iteration.  Unlike the smooth test problems considered earlier, the likelihood surface in this test was highly non-monotonic and multimodal; in this case, increasing the number of live points enough to adequately sample every mode present at a given log likelihood is unfeasible, but decreasing the sampling efficiency is tractable. Since the noise introduced in this test is a perturbation overtop a general trend, once MultiNest's bounding hyperellipsoids are enlarged sufficiently at each stage they are able to draw much less biased samples. 

\section{Strategies for Improvement}\label{sec:improve}
For problems like those analyzed in the present study, which require only a modest personal computer, a sound strategy, besides conducting rigorous convergence tests, is to use a more robust nested sampling tool. A number of other tools seem to meet the criterion of being less prone to biases than MultiNest, including POLYCHORD \citep[][]{2015MNRAS.450L..61H,2023MNRAS.521.1184L}, JAXNS \citep{2020arXiv201215286A}, dynesty \citep{2020MNRAS.493.3132S}, and UltraNest \citep{2021JOSS....6.3001B}.\footnote{Empirical evidence points to slice sampling implementations being less prone to producing biased evidence estimates than the rejection sampling-based MultiNest  (see, for example, \citet{2023MNRAS.521.1184L} and Figure 4 of \citet{2020arXiv201215286A}, although in that case other parameters must be tested to ensure convergence); other packages such as UltraNest implement more robust bounding regions for rejection sampling \citep[e.g.,][]{2016S&C....26..383B,2021JOSS....6.3001B}. Furthermore, packages such as dynesty and dyPolyChord implement dynamic nested sampling \citep{2019S&C....29..891H}, which can target both posterior and evidence accuracy in addition to providing the capability of further refining an initial nested sampling analysis; UltraNest also allows further refinements following an initial run, and targeting both posterior and evidence accuracy.} \emph{If} a given likelihood function is compatible with the JAX framework \citep{jax2018github}, JAXNS might be particularly favorable, and significantly faster than the alternatives \citep{2020arXiv201215286A}.\footnote{UltraNest also provides vectorized step samplers that can take advantage of JAX-based or otherwise GPU-parallelized likelihoods.} Otherwise, MultiNest can remain an order of magnitude or so faster than more robust sampling algorithms. As shown above, one might need to improve the resolution (and possibly expense) of a fiducial MultiNest run by orders of magnitude for the results to be worth trusting, but both this prospect and that of using a slower sampling strategy can be  intractable when committing thousands or millions of core hours to a Bayesian analysis on a supercomputer.

Generally speaking, it is possible to use an initial (presumably \emph{relatively} quick) MultiNest analysis to accelerate other methods more capable of accurately estimating evidences and posteriors. One approach is to use the initial nested sampling analysis to initialize a Markov-chain Monte Carlo (MCMC) analysis \citep[e.g.,][]{2019ApJ...887L..24M,2021ApJ...918L..28M,2023PhRvD.107j4056W}, where as long as the initial nested sampling analysis was able to correctly identify the most significant mode within a given parameter space the MCMC analysis can be expected to draw unbiased samples from the posterior. If the initial nested sampling analysis was able to adequately sample the posterior distribution, the MCMC sampler will be able to start drawing unbiased samples almost instantly. However, if the nested sampling algorithm does a poor job of estimating posteriors this method can require a long ``burn-in'' period, and standard MCMC techniques can not reliably estimate the evidence.\footnote{Parallel-tempered MCMC samples may be used to estimate the evidence as well \citep[e.g.,][]{c15dad70-7d53-348c-b458-bdfc50c4f32d,2016MNRAS.455.1919V}, but estimates using only samples of the posterior typically have \textit{infinite} variance \citep{https://doi.org/10.1111/j.2517-6161.1994.tb01956.x} and are therefore useless. See \citet{2023PhRvD.107j4056W} for a description of how samples from a single nested sampling analysis can be used to initialize a parallel-tempered MCMC analysis.} Another related approach, although I have not yet tested it personally, could be to initialize a population Monte Carlo analysis \citep[e.g.,][]{doi:10.1198/106186004X12803,Cappé2008}, similar to the description in \citet{2013arXiv1304.7808B}, which could also produce evidence estimates via importance sampling. 

Another strategy is to use a preliminary MultiNest analysis to jump-start a nested sampling analysis with a slower but more robust nested sampling strategy. Qualitatively, information about the posterior can be used to reduce the time spent by a nested sampling analysis compressing from the prior to the posterior \citep{2022arXiv221201760P}. This strategy can be used to refine a dubious initial analysis, or to accelerate the analysis a problem very similar to one already analyzed (for example, one adding a few new observations to an existing data set). Here, I have used the implementation of ``warm starting'' built into UltraNest.\footnote{\url{https://johannesbuchner.github.io/UltraNest/example-warmstart.html}. One must take care when starting from MultiNest outputs, which have a slightly different format that UltraNest outputs.}  

\begin{figure}
 \includegraphics[width=\linewidth]{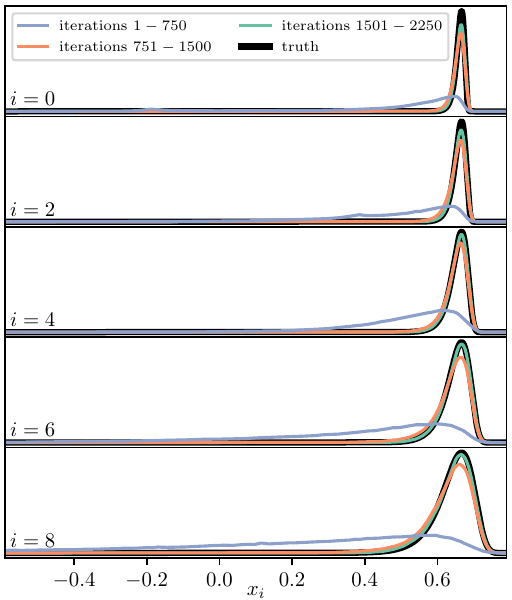}
 \caption{Probability distributions for the even parameters in a 10-dimensional log-gamma test over the course of an MCMC analysis. Initial walkers positions were sampled from the posteriors derived by an $N_{\rm live}=1000,\,\epsilon=0.1$ MultiNest analysis. Gradually, as the MCMC algorithm iterates, the walkers move from their initial positions, far from representative of the posterior, to being consistent with the underlying distributions. The distributions are plotted here using averaged shifted histograms \citep{scott1985}. 
 }
 \label{fig:mcmcfix}
\end{figure}

\begin{figure}
 \includegraphics[width=\linewidth]{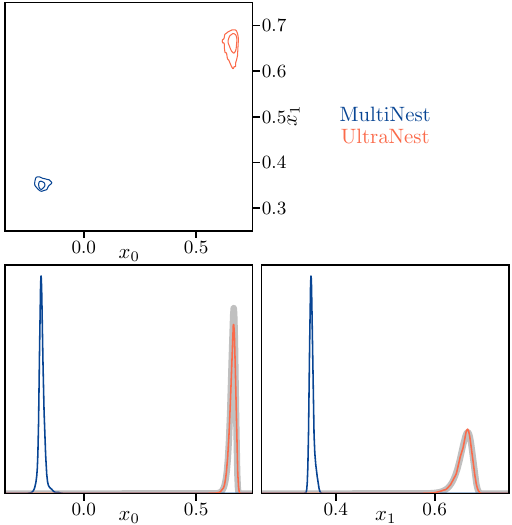}
 \caption{Results from a 10-dimensional log-gamma test, illustrated using a corner plot for the first two parameters. MultiNest results are shown in blue, while UltraNest results (warm-started from the MultiNest results) are shown in orange. The true 1D posteriors are shown in gray. The upper panel displays the two-dimensional $1\sigma$ and $2\sigma$ minimum-area credible regions. The distributions are plotted here using averaged shifted histograms \citep{scott1985}.}\label{fig:ultracomp}
\end{figure}

\subsection{MCMC}\label{sec:mcmcimprove}
For this test, I began with arguably the most difficult problem --- the 10D log-gamma distribution --- and started with a MultiNest analysis that had used inadequate hyperparameters, one of the $N_{\rm live}=1000,\,\epsilon=0.1$ runs. I first performed a Gaussian kernel density estimate of the MultiNest posterior samples; at this stage I prefer to use a bandwidth an order of magnitude or two (in this case a factor of 10) smaller than the Silverman suggestion \citep{1986desd.book.....S} to help prevent the kernel density estimate from extending beyond the prior bounds, if applicable. After estimating the density of the posterior, I resampled it, drawing 1024 samples, which I used to initialize the walkers in an MCMC analysis. For this I employed emcee version 3.1.4, using the default settings \citep{2013PASP..125..306F}. 

Some of the results from this test (for every other parameter) are shown in Figure \ref{fig:mcmcfix}. In this Figure, I have plotted the probability distributions derived from the first, second, and third groups of 750 MCMC iterations, after initializing the walkers based on draws from the MultiNest-derived posteriors. Over time, the walkers move from their initial positions to instead match the posterior. This sort of behavior can be expected as long as MultiNest is able to identify the most prominent mode in a multimodal parameter space.  This procedure also necessitates diligence during the MCMC stage to ensure convergence; as an example, if this analysis had used too few walkers, a less-efficient MCMC proposal, or fewer iterations, accurate posteriors would not have been recovered.
Because selecting good initial walker positions can be one of the greatest challenges in an MCMC analysis, these two methods can complement each other nicely. 

\subsection{Starting Warmly}
For this test, I began with the same MultiNest outputs as Section \ref{sec:mcmcimprove}, but I instead used it to warm-start an UltraNest analysis. This technique has already been demonstrated for cases where the information used to warm-start a given nested sampling run is representative of the underlying posterior \citep[e.g.][]{2022arXiv221201760P}. In this sense, the present test, where the nested sampling run used to warm-start the analysis did not accurately measure the posterior, is something like a worst-case scenario, and demonstrates the robustness of this strategy when the deviations between the warm start and actual posterior are large.

Despite using suboptimal samples to inform the warm-starting procedure, UltraNest was still able to perform adequately without any tuning. For example, UltraNest derived an evidence estimate with an error of $\Delta \log{Z}=0.690 \pm 0.605$. On the other hand, MultiNest had an evidence error of $\Delta \log{Z}=-297.245 \pm 0.259$. Additionally, as illustrated in Figure \ref{fig:ultracomp}, UltraNest was able to  in this case derive more accurate posterior estimates.  Although the aforementioned UltraNest results are encouraging, this procedure should not be taken for granted, and one still ought to perform convergence tests as part of practical analyses.

\section{Conclusions}
The results presented here imply a few general strategies for using MultiNest in a way that can be reasonably trusted. If one finds consistent evidence estimates and posterior inferences between two or more MultiNest analyses that have varied the number of live points and sampling efficiencies by factors of $\sim$three or more (larger factors are preferred because of the often-large variance between MultiNest analyses using a given combination of sampling parameters), the results of the analyses can probably be trusted. In practice, lowering the sampling efficiency may be a less resource-intensive strategy, particularly if one finds a fraction of accepted proposals higher than the sampling efficiency. Alternatively, it is straightforward to use a preliminary MultiNest run to initialize and accelerate an MCMC analysis, or another nested sampling analysis with more rigorous convergence criteria, as described in Section \ref{sec:improve}. 

To reiterate some of the results of the above tests, when used with the ``recommended'' settings, MultiNest:
\begin{itemize}
\item can produce systematically biased estimates of the Bayesian evidence (more significantly for higher-dimension problems),
\item can derive completely erroneous posterior estimates
\end{itemize}
The degree to which MultiNest produces biased results (or whether it instead produces accurate results) depends sensitively on the dimensionality and structure of a given problem. Even with relatively few ($\sim100$) live points, MultiNest can still derive accurate evidence estimates and posteriors given a sampling efficiency value low enough so that at each stage it draws samples from within the correct isolikelihood surface. MultiNest naturally experiences more difficulty sampling from likelihoods that are more difficult to approximate using ellipsoids (such as in Sections \ref{sec:bananas} and \ref{sec:logGammaTests}); since this is rarely known a priori, testing convergence is a necessary part of any analysis.

Additionally, when sampling noisy likelihoods, MultiNest tends to produce posterior estimates that are systematically narrow; this bias can be ameliorated by decreasing the sampling efficiency, but is largely independent of the number of live points used. Although the particular test problem examined here is not necessarily analogous to  any practical application, this behavior mimics that observed in some actual analyses where MultiNest seemed to significantly underestimate the widths of credible regions \citep[e.g.,][]{2019ApJ...887L..24M,2021AJ....162..237I,2021ApJ...918L..28M}. Because MultiNest's posterior estimates can be \textit{systematically} biased at larger sampling efficiencies, it is crucial that any analysis relying on MultiNest-derived posteriors test convergence with respect to the sampling efficiency,  and not only the number of live points. 
\pagebreak
\section*{Software}
MultiNest \citep{2009MNRAS.398.1601F}, PyMultiNest \citep{2016ascl.soft06005B}, emcee \citep{2013PASP..125..306F},  UltraNest \citep{2021JOSS....6.3001B},  matplotlib \citep{4160265}, numpy \citep{5725236}, scipy \citep{2020SciPy-NMeth}

\section*{Acknowledgments}
This work benefited from stimulating discussions with and useful feedback from Arjun Savel, Jegug Ih, Matt Nixon, and Cole Miller.
I am grateful for the detailed feedback and comments provided by Johannes Buchner on an draft of this manuscript. 
This work was supported in part by NASA ADAP grants 80NSSC21K0649 and 80NSSC20K0288.  I thank the reviewers for their numerous comments that led to substantial improvements of this paper.
Many of these calculations were performed on the ASTRA computing cluster, which is maintained by the Department of Astronomy at the University of Maryland. 
I also used arXiv and NASA’s Astrophysics Data System Bibliographic Services extensively.

\bibliographystyle{aasjournal}
\bibliography{references}

\end{document}